\documentclass[nofootinbib,aps,preprint,showpacs,superscriptaddress, groupedaddress]{revtex4-1}  
\usepackage{latexsym}
\usepackage{amsmath}
\usepackage[utf8x]{inputenc}
\usepackage[T1]{fontenc}
\usepackage{subdepth}
\usepackage{color}
\usepackage{soul}

\begin{document}
\def\calr{{\cal R}}
\title{On Planetary Orbits in  Entropic Gravity.}
\author{G. P\'erez-Cu\'ellar$^1$}
\email{perezcg2015@licifug.ugto.mx}
\author{M. Sabido$^{1}$}
\email{msabido@fisica.ugto.mx}
\affiliation{
$^{1}$Departamento  de F\'{\i}sica de la Universidad de Guanajuato, A.P. E-143, C.P. 37150, Le\'on, Guanajuato, M\'exico.
 }%
\begin{abstract}
Starting with an entropy  that includes  volumetric, area and length terms as well as logarithmic contributions, we derive the corresponding modified  Newtonian gravity  and derive the expression for planetary orbits. We calculate the shift of the perihelion of Mercury to find bounds to the parameters associated to the modified Newtonian gravity. We compare the parameter associated to the volumetric contribution in the entropy-area relationship with the value derived for galactic rotation curves and the value obtained from the cosmological constant.

 \end{abstract}
\maketitle

\section{Introduction}
One of the greatest achievements of theoretical physics, was the development of General Relativity (GR). With the 
LIGO detection of  gravitational waves and  the black hole shadow detection by the Event Horizon Telescope Collaboration, these remaining predictions of GR were verified. When we combine GR and the Standard Model of particle physics to study the Universe, there are two features that  have eluded explanation,  the apparent lack of matter known as the Dark Matter problem and the current accelerated expansion of the Universe, known as the dark energy problem. The mainstream approach to explain dark energy and dark matter, assumes the correctness of GR  and modifies the energy momentum tensor, by  introducing non standard sources of matter and energy (i.e, a cosmological constant, weakly interacting particle, scalar fields, among others), but current observations do not discard the possibility of using a different theory of gravity with standard matter to explain the same problems. We can make a more compelling case for the second approach when we consider that after a century of work the quantization of GR has not  produced a complete quantum theory of gravity (although there are great advances in this direction).
 There are several modified theories of gravity, $f(R)$ \cite{Sotiriou:2008rp}, massive gravity \cite{deRham:2014zqa}, Horndeski \cite{Horndeski}. All of these theories assume that gravity is a fundamental interaction. We can follow a different  approach, that gravity is not fundamental but an emergent interaction. The derivation of GR from the entropy area relationship was done in \cite{Jacobson:1995ab}, but during the last decade, there was revival in considering gravity as an emergent phenomena. This renewed interest started with the proposal that Newtonian gravity is an entropic force \cite{Verlinde:2010hp}, similar to entropic forces that are present in polymer physics. For this proposal of Newtonian gravity, the force equation is derived from the Bekenstein-Hawking entropy and therefore  we can introduce  modifications to force equation by using new entropy area relationships. More recently the entropic ideas where used to explore  a common origin for Dark Matter  and Dark Energy \cite{Verlinde2}. The predictions of this theory for the dark matter problem have been tested in several works \cite{niz,test2}. More recently the effects of Verlinde's gravity have been explored at the Solar System scale \cite{Yoon:2020xic}. Unfortunately there are some theoretical problems in the latest formulation of Verlinde, nonetheless  the entropic or emergent origin of gravity and its possible relation to  Dark Matter and Dark Energy, is a novel way to tackle these problems. In recent works \cite{isaac,isaac_ijmpd} this possibility was explored (although dark matter and dark energy where treated independently), by considering that gravity can be derived from the entropy.
 
In this work we  use the modified entropy area relation that includes the usual area term, but also volume and length contributions. {These contributions to the entropy can be obtained  from the supersymmetric generalization of the Wheeler-DeWitt (WDW) equation for Schwarzschild black holes \cite{isaac} (see Appendix \ref{apendice} for a  review of the derivation)}.  This new entropy was used to study the anomalous galactic rotation curves \cite{isaac} and in \cite{isaac_ijmpd}  to explain the origin of the cosmological constant. These two works discarded  the logarithmic and linear terms on the area, because their contribution at the galactic and cosmological scale is negligible. They concluded that the volumetric contribution to the entropy can account for the anomalous galactic rotation curve or for the late time acceleration of the Universe. For this work we will analyze the contributions of the modified entropy area relationship at the scale of the Solar System. We calculate the the planetary orbits and focus our attention on the Mercury perihelion shift  to find bounds on the entropy parameters. We find that at this scale, the logarithmic and the linear contributions  are negligible and that the parameter associated to the volumetric term has an upper bound  of order $10^{-58}$,  in accordance with the results in \cite{isaac}. \\
{This work is arranged as follows:} in section \ref{eg} we derive the modified Newtonian gravity, from an entropy that includes corrections to the Hawking-Bekenstein relationship.   We also  solve the Kepler problem and using the data for Mercury, we calculate the perihelion  shift to infer the value of the parameters in the proposed entropy.
 Section \ref{conclusiones} is devoted to  conclusions and final remarks. Finally, in Appendix \ref{apendice}  we review the derivation of the the entropy area-relationship.

\section{Planetary orbits from entropic gravity}\label{eg}
{For the entropic formulation of gravity, the main ingredient is the modified entropy area relationship and therefore one usually proposes an  expression that contains the terms we wish to explore. Therefore, let us start with the modified  entropy

\begin{equation}\label{entr}
\frac{S(A)}{k_B}=
\frac{A}{4l_{P}^{2}}+\alpha\ln{\frac{A}{4l_{P}^{2}}}
+\gamma\left(\frac{A}{4l_{P}^{2}}\right)^{1/2}+\epsilon\left(\frac{A}{2l_{P}^{2}}\right)^{3/2}.
\end{equation}
We will treat the parameters $\alpha$, $\gamma$ and $\epsilon$ as independent, this will permit us to analyze the contribution of each term independently of the origin of the entropy. Nonetheless, if we want to interpret the modification in the context of the SUSY generalization \cite{isaac}, we only need to replace the independent parameters in the entropy  with those of Eq.(\ref{parameters}) of  Appendix \ref{apendice} . 

{We can see in Eq.(\ref{entr}) not only the Hawking-Bekenstein entropy, we also have the term $A^{3/2}$. We will refer to this term as the volumetric term.  This  term is present when one considers the degrees of freedom of quantum field theories,  and is usually not  present in classical  theories of gravity. Although uncommon in gravity,  in Loop Quantum Gravity \cite{Livine}  volumetric corrections to the entropy appear in the study of black holes . There is a term proportional to $A^{1/2}$ (we will refer to this term as the length term), it describes a system with an entropy proportional to its characteristic length \cite{deVega:2001zk}. This term appears in the study  of a self-gravitating gas. Finally, the logarithmic correction  to the entropy is  has been obtained in different approaches to quantum black holes \cite{Obregon:2000zd,Domagala:2004jt,Mukherji:2002de,Sen:2012dw}.}\\\\
Although  the original derivation of Einstein's equations from the Hawking-Bekenstein entropy was done by Jacobson \cite{Jacobson:1995ab}, it was Verlinde who derived Newtonian gravity as an entropic force
\begin{equation}
F\Delta x=T \Delta S,
\end{equation}
and explicitly obtained the gravitational force from the fundamental entropy area relationship. This motivated a new approach to introduce modifications to Newtonian gravity. For example, one can introduce new terms   by considering quantum contributions to the entropy  \cite{Modesto:2010rm}, another approach is to use  alternative entropies \cite{Martinez-Merino:2017xzn} that will induce corrections to the gravitational force. These entropic corrections and their corresponding  modifications to gravity have been considered in cosmology, to derive the Friedman equations\cite{Sheykhi:2010yq}.

To obtain the general expression for the gravitational force, we assume a relationship between the entropy with the information inside  a surface {$\Omega$} that envelopes a mass $M$ and is near a mass $m$, this gives the  expression for Newtonian gravity in terms of the entropy \cite{Verlinde:2010hp} 
\begin{equation}\label{verlinde}
\mathbf{F}=-\frac{GMm}{R^{2}}\left[1+4l_{P}^{2}\frac{\partial \mathfrak{s} }{\partial A}\right]_{A=4\pi R^{2}}\mathbf{\hat{R}},
\end{equation}
where the entropy is written in terms  of $A$,  the area of the horizon. This entropy includes the Hawking-Bekenstein entropy {plus a function} $\mathfrak{s}(A)$ that encodes the corrections 
 
\begin{equation}
\frac{S}{k_{B}}=\frac{A}{4l_{P}^{2}}+\mathfrak{s}(A).\label{moden}
\end{equation}
The derivation of the modified Newtonian force is obtained from Eq.(\ref{entr}) and Eq.(\ref{verlinde}) 
\begin{equation}
\mathbf{F}_{M}=-\frac{GMm}{R^{2}}\left[1+\frac{3\sqrt{2\pi}  }{l_{P}} \epsilon R+\frac{l_{P}}{2\sqrt{\pi}}\frac{\gamma}{R}
-\frac{l_{P}^{2}}{2\pi}\frac{\alpha}{R^2}\right]\mathbf{\hat{R}}. \label{modforce}
\end{equation}  
This equation was used in the study of galaxies. For the  circular motion of a star (see \cite{isaac}),  we get  a constant velocity
\begin{equation}\label{galaxia}
v^2\approx\frac{3\sqrt{2\pi}G M}{l_P}\epsilon,  
\end{equation}
this expression is for stars far from the galactic center.  The dominating term for large $R$ is the volumetric contribution to the entropy. { To relate the correction to Newtonian gravity
with galactic rotation curves,  we compare Eq.(\ref{galaxia}) to the velocity obtained from MOND, $v^2=\sqrt{G Ma_0}$ and get
\begin{equation}\label{eps}
\epsilon=\frac{1}{6\pi}\sqrt{\frac{a_0h}{M c^3}},
\end{equation}
where  the constant $a_0$ is the  characteristic acceleration of MOND. This constant has an empirical value\cite{lotr} $a_0=1.2\times 10^{-10}m/s^2$ and considering the mass of the galaxy \cite{galaxia} $M=1\times10^{11} M_\odot$ we get\footnote{The obtained value for $\epsilon$ is an approximation. For a more robust result one should follow the approach used to constraint MOND \cite{galactic_mond}, where Solar System and galactic rotation curves constraints are combined.}    $\epsilon=6.4\times10^{-57}$. From Eq.(\ref{galaxia}), we see that the  mass $M$ scales as $v^2$ and will not satisfy the baryonic Tully-Fisher relation\footnote{The data is consistent with\ $M\propto v^s$ with $s$ between 3.5 and 4.}, thus the model is inconsistent with galactic rotation curve data\cite{Lelli:2019igz}. }

We ask ourselves, if for the Solar System the corrections derived from the modified entropy are compatible with the results at the galactic scale. 

Let us start by writing the force equation in spherical coordinates
\begin{equation}
  (\ddot{R} - R\dot{\phi}^2)\hat{R} + \frac{1}{R}\frac{d}{dt}(R^2\dot{\phi})\hat{\phi} = 
  -\frac{GM}{R^2}\left[1 + \frac{3\sqrt{2\pi}R}{lp}\epsilon + \frac{lp}{2\sqrt{\pi}R}\gamma - \frac{lp^2}{2\pi R^2}\alpha\right]\hat{R}.
    \label{eq:2N}
\end{equation}
{ Although  the contribution associated to  the logarithmic and lenght terms  are irrelevant  at large distance, we are considering these terms  for completeness.}

As in the usual Kepler problem the angular momentum is conserved and the orbits are restricted to a plane, so we identify $R^2\dot{\phi}=h$ with the magnitude of the angular momentum. Then
for radial equation we have
\begin{equation}
    \ddot{R} - R\dot{\phi}^2 = -\frac{\mu}{R^2}\left[1 + \frac{3\sqrt{2\pi}R}{lp}\epsilon + \frac{lp}{2\sqrt{\pi}R}\gamma - \frac{lp^2}{2\pi R^2}\alpha\right],
    \label{eq:dotR}
\end{equation}
where we have defined $\mu \equiv GM$.\\
Following the usual procedure for the Kepler problem we use the change of variable $u=R^{-1}$. In terms of the new variable in Eq.(\ref{eq:dotR}) we get
\begin{equation}
\frac{d^2u}{d\phi^2} + u = \frac{\mu}{h^2}\left[1 + \frac{3\sqrt{2\pi}}{l_P}\epsilon u^{-1} + \frac{l_P}{2\sqrt{\pi}}\gamma u - \frac{l_P^2}{\pi}\alpha u^2\right].
\end{equation}
Taking the parameter $\varepsilon \equiv 3\frac{\mu^2}{h^2}<<1$ 
and expanding in terms of $\varepsilon$ we get
\begin{equation}
  \frac{d^2u}{d\phi^2} + \left(1 - \varepsilon\frac{l_P}{6\mu \sqrt{\pi}}\gamma\right)u = 
  \frac{\mu}{h^2}\left[1 + \varepsilon \left(\frac{h^2\sqrt{2\pi}}{\mu^2 l_P}\epsilon u^{-1} - \frac{h^2l_P^2}{6\mu^2\pi}\alpha u^2\right)\right].
\end{equation}
Now we define $\rho \equiv \left(1 - \varepsilon\frac{l_P}{6\mu \sqrt{\pi}}\gamma\right)$ and arrive to
\begin{equation}
   \frac{d^2u}{d\phi^2} + \rho u = \frac{\mu}{h^2}\left[1 + \varepsilon \left(\frac{h^2\sqrt{2\pi}}{\mu^2 l_P}\epsilon u^{-1} - \frac{h^2l_P^2}{6\mu^2\pi}\alpha u^2\right)\right].
   \label{eq:diff}
\end{equation}

To solve this equation we will use a perturbative method, we start by assuming that the solution is given in powers of $\varepsilon$
\begin{equation}
    u = u_0 + \varepsilon u_1 +  \mathcal{O}(\varepsilon^2),
    \label{eq:prop}
\end{equation}
substituting in Eq.(\ref{eq:diff}) {and writing derivatives with respect to $\phi$ as $u'$}, we get the equations
\begin{equation}\label{eqs}
    u_0'' + \rho u_0 = \frac{\mu}{h^2},\quad
    u_1'' + \rho u_1 = \frac{ \mu}{h^2}\left(\frac{h^2\sqrt{2\pi}}{\mu^2 l_P}\epsilon u_0^{-1} - \frac{h^2l_P^2}{6\mu^2\pi}\alpha u_0^2\right).
\end{equation}
The first equation has the solution 
\begin{equation}
    u_0 = \frac{\mu}{\rho h^2}\left[1 + e\rho cos(\sqrt{\rho}\phi)\right],
    \label{eq:base}
\end{equation}
for  $\rho=1$ we get the usual Kepler problem and therefore $e=\frac{h^2}{\mu}$ is eccentricity. Introducing Eq.(\ref{eq:base})  in Eq.(\ref{eqs}) we get
\begin{eqnarray}
    &&\frac{d^2u_1}{d\phi^2} + \rho u_1=  \frac{\mu}{h^2} \left[ \frac{h^2\sqrt{2\pi}}{\mu ^2l_P}\epsilon\frac{\rho h^2}{\mu}(1 - e\rho \cos{\sqrt{\rho}\phi})\right.\\
            &&\left .-\frac{h^2l_P^2}{6\mu ^2\pi}\alpha\left(\frac{\mu}{\rho h^2}\right)^2(1 + e^2\rho^2cos^2\sqrt{\rho}\phi
        + 2e\rho cos\sqrt{\rho}\phi) \right],\nonumber
\end{eqnarray}
where we have used the approximation $e\rho<1$, from {Eq.(\ref{eq:base})} we see that this limits us to elliptic orbits. The solution for $u_1$ is
\begin{eqnarray}
    u_1 &=& \frac{\mu}{\rho h^2}\left[\frac{\sqrt{2\pi}h^4}{\mu^3l_P}\epsilon\rho - \frac{l_P^2}{6\pi h^2}\frac{\alpha}{\rho^2}\left(1 + \frac{e^2\rho ^2}{2}\right)\right.\\
    &-& \left(\frac{\sqrt{2\pi}h^4}{2\mu^3l_P}\epsilon\rho^2 + \frac{l_P^2}{6\pi h^2}\frac{\alpha}{\rho}\right)e\sqrt{\rho}\phi \sin\sqrt{\rho}\phi+\left. \frac{l_P^2}{36\pi h^2}\alpha e^2 \cos{\sqrt{4\rho}\phi}\right].\nonumber
\end{eqnarray}
{The last term is small and will be discarded, the contribution of the second term is cumulative for every revolution, therefore we will keep the first and third terms. {Defining}}
\begin{equation}
\phi ' \equiv \sqrt{\rho}\phi, \quad \delta \equiv - \frac{\varepsilon}{\rho}\left(\frac{\sqrt{2\pi}h^4}{2\mu^3l_P}\epsilon\rho^2 + \frac{l_P^2}{6\pi h^2}\frac{\alpha}{\rho}\right),
\end{equation}
 we can rewrite the solution as
\begin{equation}
    u = \frac{\mu}{\rho h^2}\left\{1 + e\rho\left[ \cos{\phi'} + \delta\phi' \sin{\phi'}\right]\right\}.
\end{equation}
Finally, taking the approximation $\delta<1$ we find
\begin{equation}
    u = \frac{\mu}{\rho h^2}\left(1 + e\rho \cos{\left[\sqrt{\rho}\phi(1-\delta)\right]}\right).
    \label{eq:eom}
\end{equation}

To calculate the precession of the orbit we calculate the angle to complete one period,
\begin{equation}
\phi_{entropic} = \frac{2\pi}{\sqrt{\rho}}(1-\delta)^{-1} \approx \frac{2\pi}{\sqrt{\rho}}(1 + \delta),
\end{equation}
which in terms of the entropy parameters gives for the precession 
\begin{equation}
     \Delta_{entropic} \approx \varepsilon 2\pi\left( \frac{ l_P\gamma}{12\mu\sqrt{\pi}} - \frac{\sqrt{2\pi}h^4}{2\mu^3l_P}\epsilon - \frac{l_P^2}{6\pi h^2}\alpha\right),
 \end{equation}
finally in terms of the orbit parameters we have
\begin{equation}\label{entropic_per}
    \Delta_{entropic} \approx \frac{\sqrt{\pi}l_P}{2a(1-e^2)}\gamma - \frac{3\pi\sqrt{2\pi}a(1-e^2)}{l_P}\epsilon - \frac{l_P^2}{a^2(1-e^2)^2}\alpha.
\end{equation}

Using the values for Mercury's orbit, 
$GM = 1.327\times10^{20} \frac{m^3}{s^2},$
 $a = 5.7909\times10^{10} m$ and
 $e = 0.2056$
 we get
  \begin{equation}
     \Delta_{{entropic}} = 2\pi\left (4.107\times10^{-47}\gamma - 1.289\times10^{46}\epsilon - 1.353\times10^{-92}\alpha\right).
 \end{equation}

Considering the the correction from the entropic modification should be small  in order to have agreement with observations, then from the previous equation we can see that $\epsilon $ should be very small.  

With this in mind, we use the reported value for orbital precession of Mercury  
$
\Delta_{obs} = 2\pi(7.98734 \pm 0.00037\times10^{-8}) \frac{rad}{rev}$, and if we consider that the entropic modifications are corrections to the prediction from $GR$,  $\Delta_{GR} = 2\pi(7.987344\times10^{-8})\frac{rad}{rev}$, 
then we have the following bound to the entropic contribution \cite{vergara}
\begin{equation}\label{delta}
 |\Delta_{entropic}| \leq |\Delta_{obs} - \Delta_{GR}| .
\end{equation}
 { Before we proceed we want to clarify that in order to arrive to Eq.(\ref{delta}), we made a strong assumption, that the entropic and the GR correction are independent. This of course is not necessarily true as there can be a cross term that includes GR and entropic effects. In order to derive the precise relation we need to derive the corrected GR equations from the new entropy area relationship following the derivation by Jacobson. Nonetheless, this approximation\footnote{This approximation was used to find a bound for the noncommutative parameter using planetary orbits [24].} can help us to determine if there is good qualitative agreement between the results from galactic rotation curves and planetary orbits.}
 
{ As already mentioned, the contributions from the logarithmic and length terms are negligible at the galactic scale, and therefore can be taken arbitrarily small to satisfy the Solar System constraints. There is also a theoretical reason to discard these term. Taking in to account the derivation of Eq.(\ref{entr}) (see appendix), from Eq.(\ref{entropy}) we see that $\alpha=1/2$ and from Eq.(\ref{parameters}) that $\gamma$ is proportional to $\epsilon$. Then discarding the $\alpha$ and $\gamma$ terms in Eq.(\ref{entropic_per}) and from $\vert\Delta_{obs}-\Delta_{GR}\vert=2.29\times10^{-11}\frac{rad}{rev}$ we get the upper bound for $\vert\epsilon\vert {\le} 2.8\times 10^{-58}$}. In the table we see the predicted values for  $\epsilon$ from Mercury and Saturn.
\begin{table}[h!]
  \begin{center}
    \label{tab:table1}
    \begin{tabular}{c | c | c} 
      \textbf{} & \textbf{Solar System} & \textbf{Galactic curve}\\
      & \textbf{limit} & \textbf{prediction} \\
      \hline
      Mercury & $\vert\epsilon\vert {\le} 2.8\times 10^{-58}$ & $\epsilon= 6.4\times 10^{-57}$\\
      \hline
      Saturn & $\vert\epsilon\vert {\le} 1.5\times 10^{-57}$&  $\epsilon= 6.4\times 10^{-57}$\\
      \hline
    \end{tabular}
  \end{center}
\end{table}

{The value of $\epsilon$ derived from galactic rotation curves 
is one order of magnitude above the upper limit  derived from the planetary constraints. }


\section{Conclusions and Final Remarks}\label{conclusiones}
One of the most interesting open problems in physics is the existence of  {\it ``dark"} components to the Universe. Does this problem originate from an incomplete formulation of gravity  or  from our ignorance on the content of the universe? The answer to this question will have a lasting effect on physics.

In this work we  study the effects from the the gravitational modifications induced by the new entropy area relationship   to the Solar System. We calculate  the effects on planetary motion and derive the expression for the perihelion shift. { The effects associated to  the logarithmic and length terms  are irrelevant at the Solar System scale unless $\alpha$ and $\gamma$ are very large, this can  be inferred from Eq.(\ref{modforce}) when considering large distances. If we consider the theoretical derivation for the entropy area relationship,  $\alpha=\frac{1}{2}$ and $\gamma\propto \epsilon$ (see Appendix) and therefore the effects of these terms are completely negligible.  The only relevant term is the modification obtained from the volumetric term  associated with the parameter $\epsilon$. From the perihelion shift of Mercury we find that the upper bound for $\epsilon$ is  $2.8\times10^{-58}$, one order of magnitude smaller than the value derived from galactic rotation curves.

We can also obtain a  prediction for $\Delta_{entropic}$ from the value of $\epsilon$ derived from Eq.(\ref{eps}). We get {\bf $\Delta_{entropic}=5.1\times10^{-10}\frac{rad}{rev}$}, the discrepancy with the value obtained from Eq.(\ref{delta}) is one order of magnitude. 
The difference between the Solar System bounds and galactic rotation seems to point that the entropic modification can not account for the anomalous galactic rotation curves. In order to be certain, we have to consider that in the calculation of the planetary bounds we assumed that the entropic and GR corrections are independent. Therefore  neglecting the possibility of an extra cross term in Eq.(\ref{delta}) that includes entropic and GR contributions, this term could account for the difference. For this we will need to use the corrected entropy to  derive the corrected GR field equations (Jacobson\cite{Jacobson:1995ab} derived Einstein's equations from the Hawking-Bekenstein entropy), and from these corrected equations calculate the perihelion shift. 
{Even if we consider the extra cross-term for the perihelion shift, it will be irrelevant for the galactic rotation curves.  These cannot be reconciled with Eq.(\ref{galaxia}) due to a wrong slope for the baryonic Tully-Fisher relation (Section 2)\cite{Lelli:2019igz}.} }

If we use  Eq.(\ref{entr}) in cosmology, we  can derive the late time limit for the Friedmann equation \cite{isaac_ijmpd}, 
\begin{equation}
H^2+\frac{\kappa}{a^2}\approx \frac{9\pi}{G}\epsilon^2.
\end{equation}
This is the  Friedmann equation 
for  de Sitter cosmology, with  the parameter $\epsilon$ playing the role of an effective  cosmological constant. Using the current value of the cosmological constant $\Lambda=1.08\times10^{-52}m^{-2}$, we find that $\epsilon$ is of order $1.8\times10^{-62}$.  Consequently, we can't simultaneously account for galactic rotation curves and late time acceleration from the modified entropy area relationship. The  bound for $\epsilon$ is  consistent with the value derived from the cosmological constant. Hence
the Solar System constraint favors an interpretation where the volumetric contribution is relevant at the cosmological scale, but not at the galaxy scale.

Finally we  mention that combining Solar System and galaxy rotation curve data is particularly powerful to discard unconventional modifications to Newtonian gravity\cite{ultimo}.


\section*{Acknowledgements}
The authors would like to thank the anonymous referee who provided valuable comments which helped to improve the manuscript. This work is supported by CONACyT grants 257919, 258982 and CIIC 290/2020. G.P.C. thanks CONACyT support.


\section{Appendix}\label{apendice}

Here we review the derivation of an entropy presented in\cite{isaac}, this entropy area-relationship contains  the usual term proportional to the area, as well as volume, length and logarithmic  contributions. This relationship is obtained from the supersymmetric WDW equation of the Schwarzschild metric \cite{Obregon:2000zd}. \\

The Schwarzschild metric is
\begin{equation}
ds^{2}=-\left(1-\frac{2M}{r}\right)dt^{2}+\left( 1-\frac{2M}
{r}\right)^{-1}dr^{2}+ r^{2}\left(d\theta^{2}+\sin^{2}\theta d\varphi^{2}\right),
\end{equation}
for  $r<2m$, the  causal structure changes. If we interchange
 $t\leftrightarrow r$ {and make the identification}
\begin{eqnarray}
e^{ -2\sqrt{3}\xi}e^{
-2\sqrt{3}\Omega}&=&t^{2},\quad
e^{2\sqrt{3}\xi}
=\frac{2M}{t}-1,\nonumber \\
N^{2}&=&\left(\frac{2M}{t}-1\right)^{-1},
\label{dif}
\end{eqnarray}
we get the Kantowski-Sachs (KS) metric 
\begin{equation}
ds^{2}-N^{2}dt^{2}+e^{\left( 2\sqrt{3}\xi\right)} dr^{2}+
e^{-2\sqrt{3}(\xi+\Omega)}  \left(
d\theta^{2}+\sin^{2}\theta d\varphi^{2}\right),
\label{KSmetric}%
\end{equation}
{this establishes the diffeomorphism  between the Schwarzschild  and KS metrics}.
The  WDW equation for the
this cosmological model is%
\begin{equation}
\left[  -\frac{\partial^{2}}{\partial\Omega^{2}}+\frac{\partial^{2}}%
{\partial\xi^{2}}+48e^{ -2\sqrt{3}\Omega } \right]
\psi(\Omega,\xi)=0.\label{ks}
\end{equation}
This result was originally derived in \cite{ro1}, where it was used to study  black holes  and determined that the wave function described Planck size states.
The supersymmetric generalization of Eq.(\ref{ks}), has been used to study  SUSY black holes \cite{Obregon:2000zd}. It is obtained by using Graham's approach \cite{Graham}, which we summarize in the following paragraphs.

The Hamiltonian $H_{0}$ for  homogeneous cosmological models  takes the form 
\begin{equation}
2H_{0}=\mathcal{G}^{\mu \nu }p_{\mu }p_{\nu }+\mathcal{U}(q^{\mu}),\label{H022}
\end{equation}
where $q^{\mu }$ and  $\mathcal{G}^{\mu \nu }$ represent the minisuperspace coordinates and minisuperspace metric. There is function $\Phi(q^{\nu})$ that satisfies 
\begin{equation}
\mathcal{G}^{\mu \nu }\frac{\partial \Phi }{\partial q^{\mu }}\frac{\partial \Phi }{\partial q^{\nu }}=\mathcal{U}(q^{\alpha}).\label{ec}
\end{equation}
The supersymmetric Hamiltonian is obtained as
\begin{eqnarray}
2H_{S}=\{\mathcal{Q},\bar{\mathcal{Q}}\},
\end{eqnarray}
where the supercharges $\mathcal{Q}$ and $\bar{\mathcal{Q}}$ are given by 
\begin{equation}
\mathcal{Q}=\psi^{\mu } \left ( p_{\mu } +i\frac{\partial \Phi }{\partial q^{\mu }}\right ),\quad \bar{\mathcal{Q}}=\bar{\psi }^{\mu }\left ( p_{\mu }-i\frac{\partial \Phi}{\partial q^{\mu }} \right ), \label{spcosm}
\end{equation}
and the Grassmann variables $\bar{\psi }^{\mu}$, and $\psi^{\nu }$  satisfy 
\begin{equation}\left \{ \bar{\psi }^{\mu } ,\bar{\psi }^{\nu } \right \}=\left \{\psi ^{\mu},\psi ^{\nu }  \right \}=0, \quad \left \{ \bar{\psi }^{\mu },\psi ^{\nu } \right \}=\mathcal{G}^{\mu \nu }.\label{alg}\end{equation}
The SUSY Hamiltonian is fully determined once we adopt a representation for $\bar{\psi }^{\mu}$, and $\psi^{\nu }$ (we have used the representation used in \cite{isaac}) such that Eq.(\ref{alg}) is satisfied.
The supersymmetric generalization of Eq.(\ref{ks}) is
\begin{equation}\label{susyec} 
\left[ -\frac{\partial^2}{\partial\Omega^2}+\frac{\partial^2}{\partial\xi^2}  + 12\left( 4\pm\frac{1}{\sqrt{e^{-2\sqrt{3}\Omega}+\epsilon^{-2/3}}} \right)e^{-2\sqrt{3}\Omega} \right] \Psi_{\pm}=0.
\end{equation}
In this equation we have  a 4 component wave function with two independent components. The supersymmetry  contributions are included  in the new potential, where the quantity $\epsilon$ is the free parameter of our model that results from the integration of Eq.(\ref{ec})  and  in the limit $\epsilon\to 0$ one recovers the original WDW equation. 

To derive the entropy we use the Feynman-Hibbs approach. In this procedure one calculates the partition function by relating the  density matrix with the kernel to the path integral. First we do a Wick rotation $t\to i\beta$, this gives is a consistent transformation from the path integral to the classical canonical partition function. 
 Now we can calculate the partition function  in a classical manner but instead of using the classical potential, we use a corrected potential (the corrected potential  includes the quantum effects).
 
Using this approach the the partition function obtained from Eq.(\ref{susyec}) is
\begin{equation}
{Z}(\beta)=\sqrt{\frac{2 \pi}{3}}\frac{1}{\beta E_{P}}\exp{\left ( -\frac{\beta^{2}E_{P}^{2}}{16\pi}-\frac{\beta^{3}E_{P}^{3}\epsilon}{96} \right )}\left ( 1+\frac{\pi \beta E_{P} \epsilon}{3} \right )^{-1/2}, \label{FPSUSY}
\end{equation}
where $\beta^{-1}=T$. From Eq.(\ref{FPSUSY}) we can directly calculate the entropy from the usual expression $\frac{S}{k_B}=\ln{{Z}}-\beta\frac{\partial}{\partial \beta}\ln {Z}$, this gives
\begin{equation}
\frac{S}{k_{B}}=-\frac{1}{2}\ln \frac{3E_{P}^{2}\beta ^{2}}{2\pi }+\frac{E_{P}^{2}\beta ^{2}}{16\pi }+\frac{E_{P}^{3}\beta ^{3}}{48}\epsilon -\frac{\pi ^{2}E_{P}^{2}\beta ^{2}}{18}\epsilon^{2}+1.
\end{equation}
Rewriting the previous expression in terms of the Bekenstein-Hawking entropy $S_{BH}/k_{B}=A/4l_P^2$, we finally arrive to 
\begin{equation}\label{entropy}
\frac{S(A)}{k_B}=\Delta(\epsilon)\frac{A}{4l_{P}^{2}}-\frac{1}{2}\ln{\frac{A}{4l_{P}^{2}}}
+\gamma(\epsilon)\left(\frac{A}{4l_{P}^{2}}\right)^{1/2}+\epsilon\left(\frac{A}{2l_{P}^{2}}\right)^{3/2},
\end{equation}
where $A$ is the area of the horizon and the functions  $\Delta(\epsilon)$ and  $\gamma(\epsilon)$
are given by
\begin{equation} \label{parameters}
\Delta(\epsilon)= 1-10\epsilon^2+4\epsilon^4,\qquad
\gamma(\epsilon)=\sqrt{2}/3(-12\epsilon+23\epsilon^3-20 \epsilon^5+4 \epsilon^7).
\end{equation}


\end{document}